\documentclass[10pt,onecolumn,peerreview,draftclsnofoot]{IEEEtran}
\usepackage{amsmath,amssymb,amsfonts} 
\usepackage{graphicx}                 
\usepackage{fontenc}
\usepackage{multirow} 
\usepackage{theorem}
\usepackage{array}

\usepackage{color,cite}
\usepackage{algorithmic}
\usepackage{algorithm}
\usepackage{epstopdf}
\usepackage{esint}
\usepackage{mathrsfs}
\usepackage[english]{babel}
\usepackage[table,xcdraw]{xcolor}
\usepackage{booktabs}
\usepackage{multirow,makecell}
\usepackage{url}

\usepackage{relsize}

\usepackage{colortbl}

\definecolor{Gray}{gray}{0.75}
\definecolor{Blue}{rgb}{0 ,0.99,0.98}
\definecolor{LightCyan}{rgb}{0.88,1,1}
\newcolumntype{b}{>{\columncolor{Gray}}c}
\newcolumntype{a}{>{\columncolor{Blue}}c}

\newcolumntype{?}{!{\vrule width 1pt}}

\usepackage{array}
\newcolumntype{P}[1]{>{\centering\arraybackslash}p{#1}}
\newcolumntype{M}[1]{>{\centering\let\newline\\\arraybackslash\hspace{0pt}}m{#1}}

\begin{document}

\title{Blockchain Technologies for Smart Energy Systems: Fundamentals, Challenges and Solutions}

\author{{Naveed UL Hassan*,~\IEEEmembership{Senior Member,~IEEE,}~ Chau Yuen,~\IEEEmembership{Senior Member,~IEEE,} and \\ Dusit Niyato,~\IEEEmembership{Fellow,~IEEE}}
\thanks{Naveed UL Hassan is with the Electrical Engineering Department, Lahore University of Management Sciences (LUMS), DHA, Lahore Cantt, 54792, Pakistan. (Email: naveed.hassan@lums.edu.pk). } 
\thanks{Chau Yuen is with the Engineering Product Development, Singapore University of Technology and Design (SUTD), 8 Somapah Road, 487372 Singapore. (Email: yuenchau@sutd.edu.sg).}
\thanks{Dusit Niyato is with the School of Computer Engineering, Nanyang Technological University, 639798, Singapore.  (e-mail: dniyato@ntu.edu.sg).}
\thanks{This research is supported by Lahore University of Management Sciences Faculty Initiative Fund (FIF) Pakistan, and partly supported by the Natural Science Foundation of China and Jiangsu Province (Project No. 61750110529, 61850410535, BK20161147).}
} 
\maketitle

\begin{abstract}
In this paper, we discuss the integration of blockchain in smart energy systems. We present various blockchain technology solutions, review important blockchain platforms, and several blockchain based smart energy projects in different smart energy domains. The majority of blockchain platforms with embedded combination of blockchain technology solutions are computing- and resource- intensive, and hence not entirely suitable for smart energy applications. We consider the requirements of smart energy systems and accordingly identify appropriate blockchain technology solutions for smart energy applications. Our analysis can help in the development of flexible blockchain platforms for smart energy systems. 

\end{abstract}

\section{Introduction}
Continuous expansion of smart energy systems for industrial, commercial and domestic applications presents several new challenges and opportunities \cite{liserre2010future,farhangi2014road}. Smart infrastructure, renewable energy sources (RES), and electrical vehicles (EVs) are becoming widespread \cite{strasser2018methods,hafez2018integrating}, energy and carbon trading possibilities are increasing \cite{sousa2019peer,tushar2018transforming,hustveit2017tradable}, and energy management (EM) through demand response management (DRM) programs are becoming more common \cite{haider2016review,ma2019demand}. In order to take full advantage of various opportunities, it becomes important to understand the requirements of smart energy systems and focus on technologies that hold the promise to fulfill those requirements. 

In recent years, there has been an increased interest in blockchain and its integration in various application domains. Blockchain is essentially a digital-distributed ledger, which is maintained and updated by a decentralized network (also called peer-to-peer (P2P) network) operating according to well-defined protocols \cite{tschorsch2016bitcoin,narayanan2016bitcoin}. A convergence of several technologies related to network, data, consensus, identity, and automation management is essential for the successful creation and implementation of a blockchain \cite{vukolic2017rethinking,xu2017taxonomy,dinh2018untangling,dunphy2018first,alharby2017blockchain}. In addition, there are also multiple technology solutions in each category. Blockchains possess several unique features, such as, decentralization, creation of a trustless network (in which nodes can resolve conflicts without a centralized authority), data storage in tamper-proof manner, fault tolerance and auditability. However, it should be noted that the choice of technology solutions has a significant impact on the resulting blockchain features and performance.
 
\begin{figure}[htb]
\centering
\includegraphics[width=0.65\textwidth]{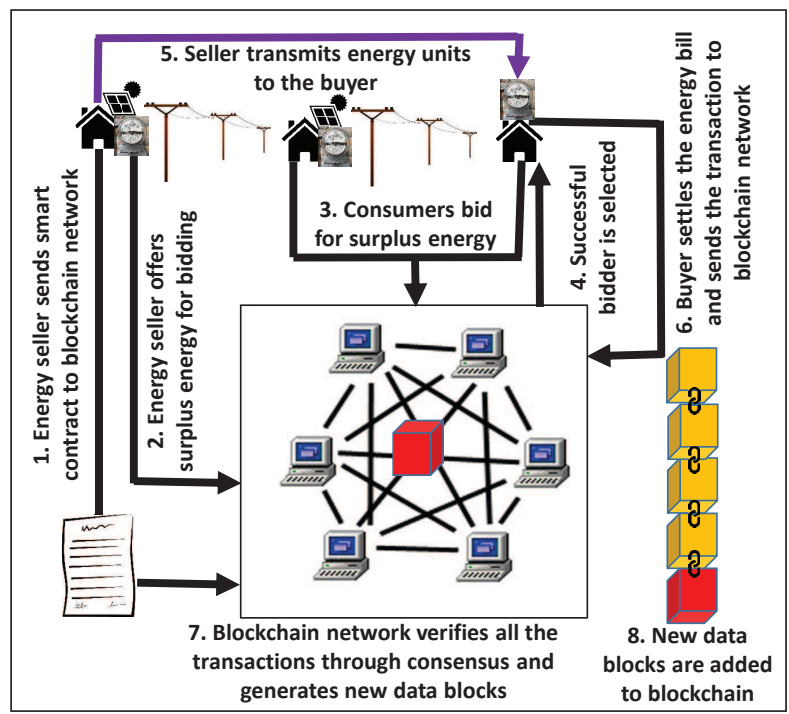}
\caption{P2P energy trading with the help of blockchain.}
\label{P2P:app}
\end{figure}

\begin{figure}[htb]
\centering
\includegraphics[width=0.8\textwidth]{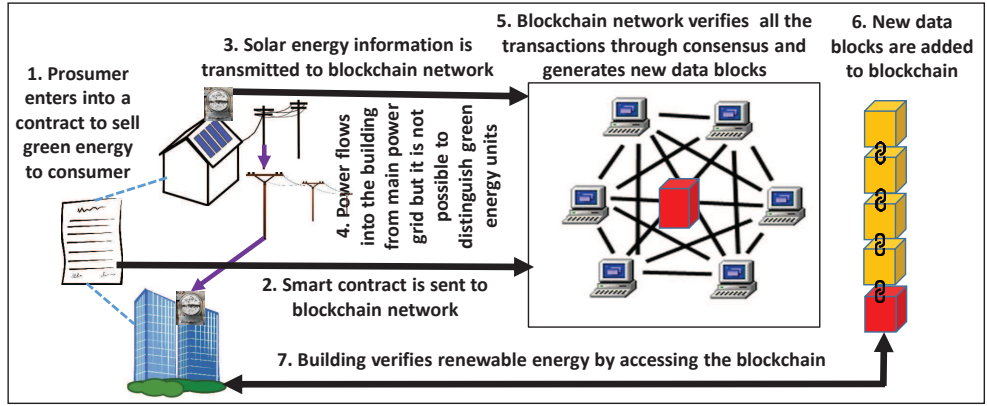}
\caption{Distributed green energy management with the help of blockchain.}
\label{green:app}
\end{figure}

The use of blockchain in smart energy systems is a topic of tremendous research interest, because further development of these systems could potentially benefit from the integration of new and innovative technologies. Blockchain due to its unique features can facilitate numerous smart energy applications. For example, Figures~\ref{P2P:app} and \ref{green:app} depict the blockchain concept and its potential role in two emerging smart energy applications. In Figure~\ref{P2P:app}, blockchain technology is being used to facilitate P2P energy trading. In this application, energy prosumers can trade surplus energy with their neighbors. However, with the introduction of blockchain, intermediaries and brokers can be eliminated because data recorded on blockchain is verified by a distributed network of nodes. Automation can be achieved through computer programs called smart contracts, which are are stored on the blockchain, and define the contractual obligations as well as the transfer of assets between peers. Another application of blockchain is shown in Figure~\ref{green:app} for the verification of green energy. Once energy is added to the grid it becomes difficult to identify green energy from the traditional energy. However, a consumer can verify the renewable energy generated by the prosumer through the use of blockchain technology. These examples demonstrate the overall concept of blockchain technology and its use in smart energy systems. However, exact blockchain technology solutions (network, data, consensus, etc.) that should converge to fulfill the requirements of these applications are not obvious. 

There are several research papers, projects, and ongoing trials that aim to leverage unique blockchain features to advance the digitalization of smart energy systems. Review of blockchain technology in energy sector can be found in \cite{andoni2019blockchain,ahl2019review,musleh2019blockchain}. In \cite{andoni2019blockchain}, authors provide a comprehensive review and classification of 140 blockchain based projects in the energy sector. In \cite{ahl2019review}, authors explore potential challenges of blockchain based P2P microgrids and they discuss a framework that incorporates technological, economic, social, environment, and institutional dimensions. The paper suggests the inclusion of economic, social, and environmental dimensions to bridge the gap between technology and institutions. In \cite{musleh2019blockchain}, authors review blockchain based smart grid projects and discuss frameworks for further blockchain integration in smart grids. According to these frameworks, creation of a cyber layer designed for blockchain applications, aggregation of computing resources in microgrids, and smart grid protection and security issues can be leveraged to achieve better integration of blockchain in smart grids.
Blockchain integration efforts in Internet of Things (IoT) are also discussed in \cite{panarello2018blockchain,ali2018applications}.
It is important to note that none of these papers identify the exact choice of blockchain technology solutions for different smart energy systems and applications.    

Blockchain technology is relatively new and although it holds tremendous potential, the number of blockchain technology solutions and implementation platforms are still developing. The choice of blockchain technology solutions that can fulfill the requirements of different smart energy applications (e.g., as shown in Figures~\ref{P2P:app} and ~\ref{green:app}) is not entirely obvious. In this paper, we provide a review of blockchain building blocks followed by the identification of the most suitable blockchain technologies according to the requirements of various smart energy systems. For example, blockchain network management techniques can be classified into public, consortium, and private categories. Similarly, data management techniques can be classified into on-chain (all data is stored on blockchain) and off-chain (only data hashes are stored on blockchain) types. Different combinations of these technology options result in different blockchain implementation platforms with different features and performance. We review important existing blockchain platforms and few representative blockchain based smart energy projects in four different domains, which include smart infrastructure (SI), energy trading (ET), green initiatives (GI), and energy management (EM). Through this review, we discover and reveal that existing blockchain platforms are not entirely suitable for smart energy systems. Therefore, in order to achieve appropriate integration of blockchain technology solutions for smart energy applications, we first consider sixteen requirements, which represent the needs of a broad selection of smart energy applications. We analyze the suitability of different blockchain technologies in fulfilling these requirements and determine appropriate blockchain building blocks for various smart energy applications. 
To summarize, our major contributions are the following:  

\begin{itemize}
\item We present a review of blockchain fundamentals and discuss various blockchain building blocks, which include network, data, consensus, identity, and automation management techniques.  

\item We review existing blockchain platforms and classify representative blockchain-based smart energy projects into SI, ET, GI, and EM domains. We show that a large number of projects use blockchain building blocks that are computing- and resource- intensive, and hence less efficient in terms of data and identity management.

\item We list 16 requirements for smart energy systems and organize them into 4 different categories, namely decentralization \& trust, data management, security, and scalability. Based on these requirements, we identify suitable blockchain building blocks that are suitable for smart energy systems and applications. 

\item We further customize blockchain technology solutions for multiple energy applications within each domain (SI, ET, GI, EM). 

\item We also identify open research areas related to blockchain technology that are needed to fulfill the future needs in smart energy systems.   

\end{itemize}

The rest of the paper is organized as follows. In Section~\ref{block:chain}, we present blockchain fundamentals and different blockchain technology solutions. In Section~\ref{block:review}, we review some blockchain integration efforts in smart energy systems. In Section~\ref{block:smart}, we identify appropriate blockchain technology solutions for various smart energy applications. In Section~\ref{block:gap}, we discuss blockchain technology gaps for smart energy integration. We conclude the paper in Section~\ref{con:clusion}.

\section{Blockchain}
\label{block:chain}
In this section, we present blockchain and various blockchain building technologies for network, data, consensus, identity, and automation management. The key points of this section are also summarized in Table~\ref{blk:features}. 
\subsection{Blockchain Fundamentals}
Blockchain is a decentralized-digital-distributed ledger. A set of transactions, which may indicate transfer or exchange of monetary value or digital assets, such as, information, services or goods is produced and collected by a distributed network of computing nodes (P2P network). A time-stamped data block (containing these transactions) is created through decentralized consensus mechanism among the nodes according to pre-defined protocols. The newly created block also contains reference to the block that came before it (parent block) in the form of cryptographic hash thus establishing a link between the blocks. The new block is added in front of its parent block and a chain like structure of blocks is obtained, hence we get the name `blockchain' (as shown in Figures~\ref{P2P:app} and \ref{green:app}). Once blockchain grows to a sufficient size, transactions recorded on it become practically immutable and resistant to change. Moreover, with blockchain, a `trustless' network of nodes is also created. In a trustless network, non-trusting nodes can interact with each other without a centralized entity or an intermediary and conflicts are automatically resolved with the help of protocols. 

\subsection{Blockchain Technology Solutions}
Blockchain creation and maintenance requires network, data, consensus, identity, and automation management. Below we present various blockchain technology options in each category and discuss their advantages and disadvantages. 
\begin{table*}[htb]
\centering
\fontsize{7.0pt}{7.0pt}\selectfont
\caption{Blockchain Technology Solutions in Different Categories with their Advantages and Disadvantages.}
\begin{tabular}{c | M{2.2cm} | M{3.6cm} | M{3.6cm} | M{3.6cm}} \hline 
\rowcolor[HTML]{9B9B9B}
\textbf{Category} & \textbf{Solutions} & \textbf{Description} & \textbf{Advantages} & \textbf{Disadvantages} \\ \hline

\multirow{3}{*}{\parbox{2cm}{\centering \textbf{Blockchain Network Management}}}
&  \textbf{Public} \textbf{(N1)} & Any node can join or leave network. & Complete decentralization with no single point of failure. & Vulnerable to Sybil attacks, high latency and less scalable.    \\ \cline{2-5}
  & \textbf{Consortium} \textbf{(N2)} & Network is controlled by a group of organizations. & More suitable for regulated industries. & Network management issues when organizations leave or join.  \\ \cline{2-5}
& \textbf{Private} \textbf{(N3)} & Network is controlled by a single organization. & More scalable, more private, cheaper to maintain. & Permission management could become a single point of failure, more centralized.  \\ \hline \hline

\multirow{2}{*}{\parbox{2cm}{\centering \textbf{Data Management}}} & \textbf{On-chain} \textbf{(D1)} & All the validated transactions are stored on the blockchain. & Greater transparency, auditability and data availability. & Huge storage burden, less scalable, not suitable for resource constrained nodes. \\ \cline{2-5}
& \textbf{Off-chain} \textbf{(D2)} & Only the hashes of important data are stored on blockchain. & Less storage requirements, suitable for resource constrained nodes. & Conventional databases required to host off-chain data. \\ \hline \hline 

\multirow{4}{*}{\parbox{2cm}{\centering \textbf{Consensus Management}}}
& \textbf{PoW} \textbf{(C1)} & Nodes compete to solve appropriate hashing puzzle. & Suitable for public networks (prevents Sybil attacks). & Wastes tremendous amount of resources, high latency, less scalable. \\ \cline{2-5}
& \textbf{PoS} \textbf{(C2)} & Nodes are picked according to their economic stake. & Suitable for public networks (prevents Sybil attacks), relatively more scalable. & Prone to ``nothing at stake" attack, less democratic. \\ \cline{2-5} 
& \textbf{Voting-based} \textbf{(C3)} & Voting schemes are based on BFT algorithms and its variants. & More suitable for consortium and private blockchain networks, low latency. & Networking and scalability issues (cannot scale beyond few hundred nodes).  \\ \cline{2-5}
& \textbf{Authority-based} \textbf{(C4)} & Trusted nodes create a new block in a round robin fashion. & Highly scalable, eliminates message exchange, more energy-efficient. & Requires trusted nodes in the network.   \\ \hline \hline

\multirow{2}{*}{\parbox{2cm}{\centering \textbf{Identity Management}}} & \textbf{Self-sovereign identity} \textbf{(S1)} & Node
owns and controls its identity without disclosure of personal data. & Guarantees more privacy. & Requires a pool of identity providers.  \\ \cline{2-5}
& \textbf{Decentralized-trusted identity} \textbf{(S2)} & Requires central server and personal data disclosures. & Establishes more trust in the network & More centralized, and less private.  \\  \hline \hline

\multirow{2}{*}{\parbox{2cm}{\centering \textbf{Automation Management}}} & \textbf{Deterministic smart contracts} \textbf{(T1)} & Does not require information from any external party. & Provides greater automation and eliminates human intervention & Execution necessitate sequential processing. \\ \cline{2-5}
& \textbf{Non-deterministic smart contracts} \textbf{(T2)} & Depends on information from an external party. & Provides more flexibility and functionality. & Non-deterministic nature, requires external party availability.  \\ \bottomrule

\end{tabular}
\label{blk:features}
\end{table*}

\noindent \underline{\textbf{Blockchain Network Management}}: 
Blockchain network management can be classified into three categories \cite{vukolic2017rethinking}.

\noindent \textbf{Public (N1)}: 
Pubic blockchain networks are truly decentralized and permissionless. Any node can join or leave the network. The nodes have full permission to maintain a complete copy of the blockchain (referred to as `public blockchain'). All the nodes can issue transactions, and they can participate in the block creation process according to publicly defined protocols and algorithms.

\noindent \textbf{Consortium (N2)}:
Consortium blockchain networks are permissioned networks. The ability of a node to join the network or to access the blockchain is controlled by a group of organizations, which assign permissions to nodes across their organizations to join the network and to read or modify the associated `consortium blockchain'. In some situations, nodes outside the consortium may also be allowed to access and read the consortium blockchain contents to achieve greater transparency. However, such nodes are not allowed to modify the blockchain state. 

\noindent \textbf{Private (N3)}:
Private blockchain networks are also permissioned networks. The network is controlled by a single organization, which allows only a limited number of nodes within the organization to join the network and to read or modify the state of `private blockchain'.  

\noindent \underline{\textbf{Data Management}}: 
Blockchain records transactions and stores data. There are two broad techniques for blockchain data management \cite{xu2017taxonomy}. 

\noindent \textbf{On-Chain (D1)}: 
In on-chain data management, all the transactions are stored on the blockchain. The size of blockchain continuously grows and storage requirements keep on increasing. This method is not suitable for resource constrained nodes. 

\noindent \textbf{Off-Chain (D2)}: 
In off-chain data management, only the hash values of data transactions are stored in the blockchain, while raw transaction data is stored using traditional methods. In this method, storage requirements at network nodes are significantly reduced. However, there are additional requirements e.g., synchronization of database with blockchain and availability of server hosting raw data. 

\noindent \underline{\textbf{Consensus Management}}:
The choice of node/nodes entrusted to create a new block depends on the consensus algorithm adopted by the blockchain network. 
Consensus algorithms allow all the nodes in the network to agree to the same world view of the state of the blockchain. There are different types of consensus algorithms \cite{dinh2018untangling,ali2018applications}.

\noindent \textbf{Proof of Work (C1)}: 
In Proof of Work (PoW) algorithm, nodes compete to solve an appropriate hashing puzzle that requires expensive computing resources. Block created by the node which is the fastest to solve the given puzzle is accepted by the network. This method is useful in permissionless networks to avoid Sybil attacks. In Sybil attack, a single node may vote multiple times with different identities to influence the vote outcome. However, PoW is energy intensive and wastes tremendous amount of resources.

\noindent \textbf{Proof of Stake (C2)}:   
In Proof of Stake (PoS) algorithm, nodes are selected to create new blocks in pseudo-random fashion. Probability of a node being selected is proportional to its economic stake in the network. This algorithm is also suitable for permissionless blockchain and punishes misbehaving nodes by confiscating their stake in the network. However, this method is prone to ``nothing at stake" attack. 

\noindent \textbf{Voting-based (C3)}:
In permissioned blockchain networks where only known nodes can join the network, consensus among validating nodes on the contents of new block can be achieved through voting mechanisms. Voting schemes are based on Byzantine fault tolerant (BFT) algorithms and its variants, such as, Tendermint and Federated BFT. In these methods multiple rounds of voting might be required to reach consensus and there is also a significant networking overhead, which has a negative impact on network scalability.  

\noindent \textbf{Authority-based (C4)}:
Proof of Authority (PoAu) algorithm can also be used in certain blockchain networks. In this mechanism, authorized (trusted) nodes in the network create a new block in a round robin fashion. PoAu eliminates message exchange among nodes for consensus building and is more resource-efficient. However, inclusion of trusted nodes reduces the trustless nature of the resulting blockchain network. 

\noindent \underline{\textbf{Identity Management}}:
Blockchain network relies on public key cryptography. Each node has a pair of public/private key to sign and verify transactions. There are different ways to manage the identity and entitlements of blockchain nodes \cite{dunphy2018first}.     

\noindent \textbf{Self-sovereign identity (S1)}:
In this method, every node owns and controls its identity without relying on an external authority for attestation or verification of node credentials. There is no central server and personal data is not required for identity creation. Nodes can perform identity proofing by gathering attributes from an ecosystem of identity providers. Each node is allowed to create multiple keys as required to keep its identity private. Nodes can also selectively disclose their attributes to maintain privacy. Sovrin and uPort are examples of self-sovereign identity management systems.
  
\noindent \textbf{Decentralized-trusted identity (S2)}:
This method requires a central server to perform identity proofing of nodes. In the initial stage, a node has to provide identity proof (personal information) to the central server. After this bootstrap phase, node identity is recorded in the blockchain for later validation. Verified nodes can then create further keys as required. ShoCard and BitID are some examples of decentralized-trusted identity management system.

\noindent \underline{\textbf{Automation Management}}:
Automation management on blockchain is carried out with the help of smart contracts, which may define contractual obligations, custody or transfer of digital assets and rights and privileges of nodes. Smart contracts provide greater automation and replicate actions that are generally performed by trusted third parties or intermediaries. Turing-complete programming languages that can support arbitrary logic and computations are generally required to develop smart contracts. We can broadly classify smart contracts into two types \cite{alharby2017blockchain}.

\noindent \textbf{Deterministic smart contracts (T1)}: 
Deterministic smart contracts do not require any information from external party. All the necessary information to execute a smart contract can be obtained from the data already stored on the blockchain. 

\noindent \textbf{Non-deterministic smart contracts (T2)}: 
Non-deterministic smart contracts depend on information (called oracles or data feeds) from an external party, e.g. it may need external weather information for execution. Non-deterministic smart contracts provides greater flexibility at the expense of greater vulnerability to external attacks. 

There is a wide variety of blockchain technology solutions. Combination of blockchain building blocks also result in different tradeoffs and different blockchain features. In addition, the requirements of different smart energy applications are also different. However, before identifying the best possible blockchain technology solutions for various smart energy applications, we first provide a brief review  of existing blockchain platforms and blockchain integration efforts in smart energy systems.

\section{Review of Blockchain Integration in Smart Energy Systems}
\label{block:review}
In this section, we review some blockchain integration efforts in smart energy systems. 
Please note that in this section, we do not intend to provide a complete survey of blockchain integration efforts in smart energy systems. A comprehensive review and classification of 140 blockchain based projects in the energy sector is available in~\cite{andoni2019blockchain}. In this section, we only present some selected platforms and projects in each smart energy domain with the objective to reveal that most of these efforts do not use blockchain technologies customized for energy applications. This review will further facilitate us in the identification of the most suitable blockchain technology solutions according to the requirements of smart energy systems. The contributions of this section are summarized in Table~\ref{tab:plat} and Figure~\ref{sg:applications}.

\subsection{Review of Blockchain Platforms Used in Smart Energy Systems}
Blockchain platforms combine network, data, consensus, identity, and automation management technologies for the creation of blockchain based projects. Blockchain integration in smart energy applications is being carried out either using open-source or proprietary blockchain platforms. Popular open-source platforms include Ethereum, HyperLedger, Tendermint, and Energy web foundation (EWF). Proprietary platforms are developed to suite the requirements of specific applications and sometimes these platforms also develop proprietary management protocols and algorithms. It should be noted that a majority of open-source and proprietary platforms are non-modular \cite{belotti:hal-01870617}. 

\noindent \underline{\textbf{Ethereum}}: 
Ethereum is a generic open source blockchain development platform governed by Ethereum developers and it is widely used for developing blockchain applications for smart energy systems \cite{ethereum_whitepaper}. This platform was developed for public (N1) blockchain management. However, the open-source code of Ethereum can be easily modified to maintain consortium (N2) and private (N3) networks. Ethereum supports on-chain data management (D1). PoW (C1) consensus algorithm is currently used but there are plans to switch to PoS (C2) algorithm. The platform can support self-sovereign (S1) as well as decentralized-trusted (S2) identity management techniques. Ethereum supports Turing-complete programming languages (Solidity and Serpent), which can be used to create deterministic (T1) as well as non-deterministic (T2) smart contracts.

\noindent \underline{\textbf{HyperLedger}}: 
HypberLedger is an open-source blockchain development platform supported by The Linux foundation \cite{hyperledger_whitepaper}. This platform can be used to set up consortium (N2) and private (N3) networks. The platform supports on-chain data management (D1), and voting-based consensus (C3) algorithms. This platform can support self-sovereign (S1) as well as decentralized trusted (S2) identity management techniques. Turing-complete programming languages such as, Java, Go, Solidity, Fabric and Rust, allow writing deterministic smart contracts (T1). However, support for non-deterministic smart contracts (T2) through oracles is not yet available. 

\noindent \underline{\textbf{Tendermint}}:
Tendermint is another application oriented framework that can be used to set up public, consortium or a private network of P2P nodes (N1,N2,N3) \cite{tendermint_whitepaper}. This platform supports on-chain data management (D1) and voting-based (C3) consensus algorithms. This platform can support self-sovereign (S1) as well as decentralized-trusted (S2) identity management techniques. The platform supports various Turing-complete programming languages that currently allows writing deterministic smart contracts (T1).  

\noindent \underline{\textbf{EWF}}:
EWF blockchain platform is supported by more than 70 companies and its aim is to integrate and accelerate blockchain technology in smart energy systems \cite{ewf_whitepaper}. EWF platform is Ethereum-compliant but it is more customized for smart energy applications. EWF platform can be used to set up consortium (N2) and private (N3) networks, supports on chain-data management (D1), and PoAu (C4) consensus algorithm. This platform can support self-sovereign (S1) as well as decentralized-trusted (S2) identity management techniques. Deterministic (T1) and non-deterministic (T2) smart contracts can be developed in Turing complete C and C++ programming languages. Tobalaba, which is the test version of this platform is already available for developers. 

\noindent \underline{\textbf{Proprietary}}:
Several proprietary blockchain platforms also exist for smart energy applications. For example, Solar Bankers is developing a proprietary consensus algorithm called Obelisk which runs on their Skychain blockchain \cite{solarbankers_whitepaper}. The idea is based on developing a trusted consortium of nodes, which generate and validate data blocks. Similarly, PROSUME is also developing a proprietary blockchain based platform to support a multitude of smart energy applications \cite{prosume_whitepaper}.

In Table~\ref{tab:plat}, we provide a summary of blockchain technology solutions supported by these platforms. 

\begin{table*}[ht]
\centering
\fontsize{7pt}{7pt}\selectfont
\caption{Review of Blockchain Platforms Used in Smart Energy Systems. Here, $\checkmark$ is used if platform supports a certain technology solution, x is used if it does not support, while - is used to represent unavailable or diverse information.}%
\begin{tabular}{|M{1.5cm}|a|a|a|b|b|a|a|a|a|b|b|a|a|M{4.5cm}|} \hline 
\multicolumn{1}{|c|}{\multirow{3}{1.5cm}{\textbf{Platform}}} &  \multicolumn{13}{c|}{\multirow{1}{*}{\textbf{Blockchain Technology Solutions}}} & \multicolumn{1}{c|}{\multirow{3}{*}{\textbf{Projects}}} \\ \cline{2-14}
& \multicolumn{3}{c|}{\textbf{Network}} & \multicolumn{2}{c|}{\textbf{Data}} & \multicolumn{4}{c|}{\textbf{Consensus}} & \multicolumn{2}{c|}{\textbf{Identity}} & \multicolumn{2}{c|}{\textbf{Automation}} & \\ \cline{2-14}
& \textbf{N1} & \textbf{N2} & \textbf{N3} & \textbf{D1} & \textbf{D2} & \textbf{C1} & \textbf{C2} & \textbf{C3} & \textbf{C4} & \textbf{S1} & \textbf{S2} & \textbf{T1} & \textbf{T2} & \\ \hline \hline

\textbf{Ethereum} & \checkmark & \checkmark & \checkmark  & \checkmark  & x  & \checkmark  & \checkmark  & x  & x  & \checkmark  & \checkmark  & \checkmark & \checkmark & Bankymoon, TheSunExchange, Brooklyn Microgrid, NRGCoin   \\ \hline

\textbf{HyperLedger} & x & \checkmark & \checkmark  & \checkmark  & x  & x  & x  & \checkmark  & x  & \checkmark  & \checkmark  & \checkmark & x & Car eWallet, Tennet \& Sonnen, SunChain    \\ \hline

\textbf{Tendermint} & \checkmark & \checkmark & \checkmark  & \checkmark  & x  & x  & x  & \checkmark  & x  & \checkmark  & \checkmark  & \checkmark & x & GridChain, EnerChain, Brooklyn Microgrid \\ \hline

\textbf{EWF} & x & \checkmark & \checkmark  & \checkmark  & x  & x  & x  & x  & \checkmark  & \checkmark  & \checkmark  & \checkmark & \checkmark & Slock.it, GridSingularity, Share\&Charge  \\ \hline

\textbf{Proprietary} & - & - & -  & -  & -  & -  & -  & -  & -  & -  & -  & - & - & Nasdaq Linq, Solar Bankers, PROSUME  \\ \hline

\end{tabular}
\label{tab:plat}
\end{table*}

\subsection{Review of Blockchain Based Smart Energy Projects}
\begin{figure}[htb]
\centering
\includegraphics[width=1.0\textwidth]{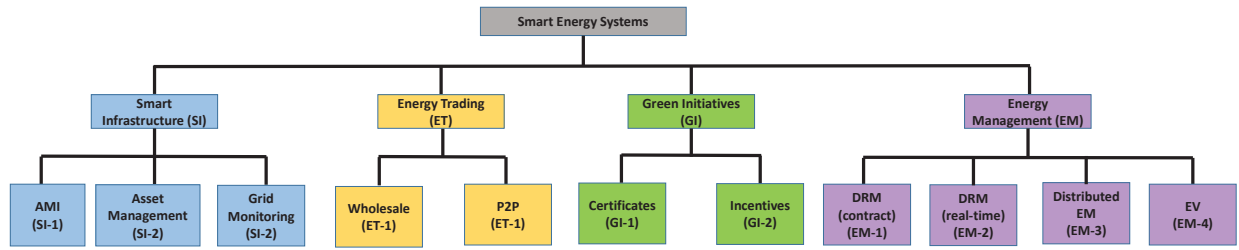}
\caption{Smart energy system domains and applications.}
\label{sg:applications}
\end{figure}

We review blockchain based smart energy projects in four smart energy domains, which include SI, ET, GI, and EM. These domains are broad and cover several interesting and useful applications. The list of domains and considered applications in each domain are presented in Figure~\ref{sg:applications}. A short notation for each application is also introduced for further use in the paper. For example, SI-1 notation is used for automated metering infrastructure (AMI) application. 
The scenario in Figure~\ref{P2P:app} represents ET-2 application, while that in Figure~\ref{green:app} represents EM-3 application.  

Due to space limitations, in the following, we only discuss few representative projects in each domain. Further details of these projects can be found in \cite{andoni2019blockchain,wu2018application,panarello2018blockchain,goranovic2017blockchain} and references therein. 

\noindent \underline{\textbf{Blockchain Projects in SI Domain}}: 

\noindent \textbf{Bankymoon}: This project is related to AMI SI-1 application. Smart meters compute and communicate energy consumption of an industrial or residential building at regular intervals for billing automation and reduction of electricity theft incidents \cite{alahakoon2015smart,kabalci2016survey}. However, in Bankymoon project, blockchain enabled smart meters are being developed and experimented in order to further automate financial transactions. These meters can be loaded with cryptocurrencies and payments can be settled in real-time through smart contracts. This project is being developed using Ethereum platform.

\noindent \textbf{TheSunExchange}: This project is related to asset management SI-2 application. High initial costs of RES technologies could become a barrier in taking communities off-grid. However, this issue may be resolved by creating shared assets in smart energy systems e.g., by purchasing solar PVs through crowd-funding \cite{lam2016crowdfunding}. TheSunExchange project allows users to purchase solar panels and lease them to earn passive income. Blockchain integration enables transparent management of assets as well as the management of solar energy produced by these assets. Therefore, this project can be also be classified as an example of example of EM-3 application, which is related to distributed EM. 

\noindent \textbf{GridChain (PONTON)}: This project is related to power grid monitoring SI-3 application. In power grids, IoT sensors in the transmission and distribution systems facilitate monitoring of grid parameters in order to automate fault diagnosis and to maintain power-balance for grid stability \cite{zidan2017fault}. Blockchain integration can further help in achieving transparency and fixing liability. In this context, the objective of GridChain project developed by PONTON is to enable real-time power balance and congestion management by providing coordination between various grid entities. This project can also be classified as an example of real-time DRM application (EM-2). 

\noindent \underline{\textbf{Blockchain Projects in ET Domain}}: 

\noindent \textbf{EnerChain (PONTON)}: This project deals with wholesale energy trading ET-1 application. The integration of blockchain in energy trading applications achieves greater transparency and automation. EnerChain project is also developed by PONTON to enable wholesale energy trading in European regional power markets. The project aims to offer wholesale energy trading solutions in different time frames such as, day-ahead, monthly, quarterly, and yearly.     

\noindent \textbf{Brooklyn Microgrid}: This project is related to P2P energy trading ET-2 application, which is shown in Figure~\ref{P2P:app}. In smart energy systems, prosumers can engage in decentralized energy trading activities where they can directly trade energy with other prosumers or consumers \cite{li2017consortium,kang2017enabling}. Brooklyn Microgrid project is an example of real-world development of blockchain based P2P energy trading solution. In this project, prosumers can directly sell their surplus energy to their neighbors (without needing any brokers or intermediaries), energy transactions are recorded on blockchain, and payments are settled automatically through smart contracts. 

\noindent \underline{\textbf{Blockchain Projects in GI Domain}}: 

\noindent \textbf{Nasdaq Linq}: This project is related to the management and trading of green certificates and carbon credits GI-1 application. To encourage RES uptake, several countries and states issue green certificates and carbon credits \cite{hustveit2017tradable}, which can also be traded. However, with greater integration of RES in power grids, management of these certificates is becoming challenging. In this context, Nasdaq Linq project aims to bring efficiency, quick verification, and elimination of paper records for green certificate management through the integration of blockchain. This project is being developed using proprietary platform.

\noindent \textbf{NRGcoin}: This project is related to the management of incentives for green behavior GI-2 application. In this project, NRGcoins are given as a reward to incentivize local production and consumption of green energy. It should be noted that 1~NRGcoin is equivalent to 1kWh energy. The use of virtual currency in this project creates additional value around their blockchain. However, unlike Bitcoin, these coins are not mined but are issued by the blockchain developers. The smart contract framework of this project is based on Ethereum platform.  

\noindent \underline{\textbf{Blockchain Projects in EM Domain}}: DRM is an important concept in smart energy grids. However, blockchain based projects for EM-1 and EM-2 applications are relatively rare. 

\noindent \textbf{Key2Energy}: This project is related to distributed energy management EM-3 application. In this project, blockchain is used for energy management in multi-apartment houses. The objective is to maximize the profit of each house by selling PV energy and minimizing the energy cost of shared facilities in the building. Platform details of this project are not available. 

\noindent \textbf{Car eWallet}: Number of EVs with batteries is increasing. Due to mobility, management of EVs and their energy consumption becomes quite challenging \cite{silva2018modern}. Car eWallet project is related to EV EM-4 application. This project provides a blockchain based solution for car sharing, car rental, and EV charging. The project also allows automatic processing of payments. 

In Table~\ref{tab:plat}, blockchain platforms used for these projects are also identified. It should be noted that several blockchain platforms (except EWF) are not exclusively developed for smart energy applications. Therefore, the embedded technology options in these platforms are also not entirely suitable for these applications. For example, a large number of projects use Ethereum, which embeds computing-intensive PoW algorithm. Similarly, several platforms lack capabilities to support off-chain data management as well as non-deterministic smart contract management. It is also important to note that most of the blockchain based smart energy projects are still in the development or trial phases, while real-world implementations are rare. In this context, in order to guide further research and development in this field, there is a need to identify appropriate blockchain technology solutions according to the requirements of smart energy systems. In the next section, we discuss these requirements and accordingly identify appropriate choice of blockchain technology solutions for various applications.


\section{Appropriate Choice of blockchain technologies according to the smart energy system requirements}
\label{block:smart}
We first discuss suitable blockchain technology solutions according to the requirements of smart energy systems followed by the customization of these solutions for various applications. We summarize the key contributions of this section in Tables~\ref{sg:blk1} and ~\ref{sg:app:blk}. 

\subsection{Suitable Blockchain Technology Solutions According to Smart Energy System Requirements}
We first discuss a total of sixteen requirements (R1-R16) in four different categories, which are applicable to a broad selection of smart energy applications listed in Figure~\ref{sg:applications} and to the scenarios depicted in Figures~\ref{P2P:app} and \ref{green:app}. 

\noindent \underline{\textbf{Decentralization \& Trust Requirements}}: 

\noindent \textbf{Decentralization (R1)}: Due to the inclusion of RES and mobile loads (EVs), the architecture of smart energy systems is becoming decentralized. Efficient implementation of various applications in different domains requires decentralized networking and control. 

\noindent \textbf{Conflict Resolution Mechanism (R2)}: Smart energy domains involve interaction between multiple non-trusting nodes. Some mechanisms (entities or technologies) are therefore required to mediate between nodes in order to resolve conflicts. 

\noindent \textbf{Intermediaries (R3)}: In several smart energy applications, intermediaries are required to support the activities of principal players. The role of intermediaries arise due to operational and technological limitations of the principal players. For example, financial transactions between consumers and generators are mostly settled through banks. Similarly, brokers or energy trading platforms are required to match the buying and selling requirements of generators and consumers. 

\noindent \textbf{Non-repudiability (R4)}: Non-repudiability refers to the availability of irrefutable proof of who performed a certain action even if the nodes are not cooperating. In smart energy domains, non-repudiability is required to establish liability.  

\noindent \underline{\textbf{Data Management Requirements}}

\noindent \textbf{Tamper-proof record keeping (R5)}: Recording, trading and transportation of electricity, assets, and other resources is involved in various smart energy systems. It is also important to note that in several situations electricity flow occurs almost immediately while financial settlements are carried out later. Therefore, it becomes important to store data in a tamper-proof manner. 

\noindent \textbf{Data correction \& erasure (R6)}: In the event of malfunction, hacking, or tampering of sensors or equipment, wrong data could get recorded. If such events are detected or reported, data correction or data erasure becomes essential. With increased automation, all the smart energy domains require a certain ability to correct and erase such erroneous data.     

\noindent \textbf{Data Backup (R7)}: Data loss can create inconvenience, disruption and financial loss. Similarly, data storage and retrieval from a single database also requires permanent availability of the data hosting node. Thus, a single point of failure is created in centralized systems. Data collected in various smart energy systems domains is often critical and important and therefore requires adequate backup to ensure smooth operations. 

\noindent \textbf{Privacy Protection (R8)}: In various smart energy systems there is a high requirement to keep data and node identity private. For example, smart meter data reveals private information about the habits, schedule and behavior of users.

\noindent \underline{\textbf{Security Requirements}}:

\noindent \textbf{Authentication (R9)}: Authentication is concerned with determining the identity of a node in the system in order to block unauthorized access. A node can be authenticated through its unique credentials in the system (e.g., public key, address, name). Smart energy systems often involve critical data and infrastructure. Therefore, authentication is always required in all the smart energy domains.  

\noindent \textbf{Authorization (R10)}: Authorization deals with managing access and privileges of various nodes in the network. In smart energy systems, nodes have different roles and therefore require different authorization in different applications. In addition, there is also a certain role of regulatory bodies and government agencies. Therefore, appropriate authorization and detecting any violations of privileges and rights is required in such systems.    

\noindent \textbf{Data Integrity (R11)}: Data integrity refers to the detection of unauthorized changes in data. Decentralized architecture requires large number of critical messages exchanged between various nodes and data integrity violations can result in safety problems or harmful attacks on the critical infrastructure. 

\noindent \textbf{Auditability (R12)}: Auditability is concerned with the ability to reconstruct complete history of certain event or action from the historical records. In smart energy systems, auditability is required to fix liability in case of malfunctions or conflicts or to safeguard commercial and financial interests or to fulfill regulatory requirements. 

\noindent \underline{\textbf{Scalability Requirements}}:

\noindent \textbf{Throughput (R13)}: In smart energy systems, a single node often produces a small amount of data. However, a large number of nodes are involved to build meaningful applications. If the data requirements of a single node is considered as a single transaction then a large number of transactions happen every second. Therefore, smart energy systems require high data throughout.

\noindent \textbf{Latency (R14)}: Smart energy applications require low latency in order to ensure smooth monitoring, control and operation of appliances, equipment and processes. Latency of some critical applications e.g., required for grid stabilization, is only few ms. 

\noindent \textbf{Process Automation (R15)}: Smart energy systems are built on the promise of making RES integration, energy transportation and energy trading more efficient. This can be achieved through increased process automation resulting in the reduction in human intervention and simplification of legacy procedures. 

\noindent \textbf{Cost (R16)}: Smart energy systems integrate novel technologies and new equipment (smart meters, sensors, etc.), which help reduce various operating costs. However, high upfront costs due to equipment replacement or technology up-gradation is a major barrier in the adoption of various concepts. In this context, all the smart energy domains can benefit from cost reductions.

Based on these requirements, we can now determine the suitability of blockchain technology solutions for various smart energy systems and applications. The suitability analysis is presented in Table~\ref{sg:blk1}. This analysis is carried out by matching the features, advantages and disadvantages of various blockchain technology solutions discussed in Section~\ref{block:chain} with smart energy system requirements. Based on this analysis, consortium (N2) and private (N3) network management emerge as more suitable options for such systems. Off-chain data management (D2) can also fulfill more requirements as compared to on-chain data management technique. Similarly authority-based consensus management (C4) is the best consensus algorithm for smart energy systems, while self-sovereign identity management (S1) and deterministic smart contracts (T1) can fulfill more requirements. This analysis enables quick identification of appropriate blockchain technology solutions for smart energy systems. However, different smart energy applications, such as, P2P energy trading and distributed green energy management (as shown in Figures~\ref{P2P:app} and ~\ref{green:app}) also have slightly different requirements. Hence, there is also a further need to customize blockchain technology solutions for various smart energy applications.  


\begin{table*}[ht]
\centering
\fontsize{6.7pt}{6.7pt}\selectfont
\caption{Suitability of Blockchain Technology Solutions in fulfilling Smart Energy Systems Requirements. Here, $\checkmark$ is used if blockchain technology is suitable, x is used if it is not suitable and - is used if it is unconcerned.}%
\begin{tabular}{|c|c|a|a|a|b|b|a|a|a|a|b|b|a|a|} \hline 
\multicolumn{1}{|c|}{\multirow{3}{1.5cm}{\textbf{Category}}} & \multicolumn{1}{c|}{\multirow{3}{*}{\textbf{Requirement}}} & \multicolumn{13}{c|}{\multirow{1}{*}{\textbf{Blockchain Technology Solutions}}} \\ \cline{3-15}
& & \multicolumn{3}{c|}{\textbf{Network}} & \multicolumn{2}{c|}{\textbf{Data}} & \multicolumn{4}{c|}{\textbf{Consensus}} & \multicolumn{2}{c|}{\textbf{Identity}} & \multicolumn{2}{c|}{\textbf{Automation}} \\ \cline{3-15}
& & \textbf{N1} & \textbf{N2} & \textbf{N3} & \textbf{D1} & \textbf{D2} & \textbf{C1} & \textbf{C2} & \textbf{C3} & \textbf{C4} & \textbf{S1} & \textbf{S2} & \textbf{T1} & \textbf{T2} \\ \hline \hline

\multirow{4}{*}{\parbox{1.8cm}{\centering \textbf{Decentralization \& Trust}}}
& \textbf{Decentralization} \textbf{(R1)}   & \checkmark & \checkmark & \checkmark  & -  & -  & \checkmark  & \checkmark  & \checkmark  & \checkmark  & \checkmark  & x  & - & -  \\\cline{2-15} 
& \textbf{Conflict Resolution Mechanism} \textbf{(R2)}      & x  & \checkmark  & \checkmark  & - & - & \checkmark  & \checkmark  & \checkmark  & \checkmark  & x  & \checkmark  & \checkmark  & \checkmark  \\\cline{2-15} 
& \textbf{Intermediaries} \textbf{(R3)}      & - & - & - & - & - & - & - & - & - & - & - & \checkmark & \checkmark  \\\cline{2-15} 
& \textbf{Non-repudiability} \textbf{(R4)}  & - & - & - & \checkmark & \checkmark & \checkmark & \checkmark & \checkmark & \checkmark & \checkmark & \checkmark & \checkmark & \checkmark  
\\\hline \hline

\multirow{4}{*}{\parbox{1.8cm}{\centering \textbf{Data Management}}} 
&  \textbf{Tamper-proof records} \textbf{(R5)}         & - & - & - & \checkmark & x & \checkmark & \checkmark & \checkmark & x & - & - & - & -  \\\cline{2-15} 
&  \textbf{Data correction \& erasure} \textbf{(R6)}   & x & \checkmark & \checkmark & - & - & x & x & x & \checkmark & - & - & - & -  \\\cline{2-15} 
&  \textbf{Data Backup} \textbf{(R7)}                  & \checkmark & \checkmark & \checkmark & \checkmark & \checkmark & - & - & - & - & - & - & - & - \\\cline{2-15} 
&  \textbf{Privacy Protection} \textbf{(R8)}                      & x & \checkmark & \checkmark & x & \checkmark & x & x & x & \checkmark & \checkmark & x & - & -  
\\\hline \hline

\multirow{4}{*}{\parbox{1.8cm}{\centering\textbf{Security}}} 
&  \textbf{Authentication} \textbf{(R9)}   & x & \checkmark & \checkmark & - & - & - & - & - & - & \checkmark & \checkmark & - & -  \\\cline{2-15} 
&  \textbf{Authorization} \textbf{(R10)}   & x & \checkmark & \checkmark & - & - & - & - & - & \checkmark & \checkmark & \checkmark & \checkmark & \checkmark  \\\cline{2-15} 
&  \textbf{Data Integrity} \textbf{(R11)}  & - & - & - & - & - & - & - & - & - & \checkmark & \checkmark & \checkmark & x  \\\cline{2-15} 
&  \textbf{Auditability} \textbf{(R12)}    & - & - & - & \checkmark & \checkmark & - & - & - & - & \checkmark & \checkmark & \checkmark & \checkmark  
\\\hline \hline

\multirow{4}{*}{\parbox{1.8cm}{\centering\textbf{Scalability}}} 
&  \textbf{Throughput} \textbf{(R13)}   & - & - & - & x & \checkmark & x & \checkmark & x & \checkmark & - & - & x & x  \\\cline{2-15} 
&  \textbf{Latency} \textbf{(R14)}      & - & - & - & \checkmark & x & x & \checkmark & x & \checkmark & - & - & - & - \\\cline{2-15} 
&  \textbf{Process Automation} \textbf{(R15)}   & \checkmark & \checkmark & \checkmark & - & - & \checkmark & \checkmark & \checkmark & \checkmark & \checkmark & \checkmark & \checkmark & \checkmark  \\\cline{2-15} 
&  \textbf{Cost} \textbf{(R16)}     & x & \checkmark & \checkmark & x & \checkmark & x & \checkmark & \checkmark & \checkmark & \checkmark & x & \checkmark & \checkmark  
\\\hline 

\end{tabular}
\label{sg:blk1}
\end{table*}

\subsection{Customization of Blockchain Technology Solutions for Various Smart Energy Applications} 
Requirements of smart energy applications differ from each other, which necessitate further customization of blockchain technology solutions. For example, some applications require low latency, some require high privacy protection, etc., \cite{gungor2013survey,yan2012survey}. In this subsection, we further identify appropriate blockchain technology solutions for various smart energy applications shown in Figure~\ref{sg:applications}. The discussion below is also summarized in Table~\ref{sg:app:blk}.

\noindent \underline{\textbf{SI domain}}: 
AMI SI-1 application has relatively relaxed latency and throughput requirements. For this application, consortium (N2) and private (N3) network management techniques can be used. Private network management is preferred if data is directly handled by the utility. Since smart meters are resource constrained nodes, off-chain data management (D2) technique is more suitable. For consensus management, PoS (C2) and PoAu (C4) algorithms are better options. This application requires high privacy protection. However, because of regulatory and registration requirements of smart meter with utility company, self-sovereign identity (S1) management cannot be used. Instead decentralized-trusted identity (S2) management is a more suitable option. Moreover, necessary automation, if required, can be managed with the help of deterministic smart contracts (T1). 
Asset management SI-2 application has relatively low privacy and throughput requirements. For this application, choice of network, data, consensus, and automation management is the same as SI-1. For managing shared RES, PoS (C2) is more suitable option. However, when there are low latency requirements, PoAu (C4) algorithm is more preferable. For identity management, self-sovereign (S1) as well as decentralized-trusted identity (S2) management are suitable. However, if Know Your Customer (KYC) requirements are not applicable then self-sovereign (S1) technique can also be used. 
Grid monitoring application has extremely stringent latency requirements (few ms). For this application, network, data, identity, and automation management options are the same as identified for SI-1 application. However, due to extremely low latency requirements, only PoAu (C4) is a suitable consensus management solution for this application. However, even this algorithm can fail to meet the required performance. \\
\noindent \underline{\textbf{ET domain}}: 
Wholesale energy trading application has relatively low privacy requirements, therefore, public and consortium network management techniques (N1,N2) are more suitable. N1 should be used if trading platform is being developed across multiple regional markets. For more localized P2P energy trading ET-2 application, only consortium network management technique (N2) is suitable. For both the applications, off-chain data management (D2) technique is more suitable. For consensus, all the options are suitable for ET-1. For ET-2, PoW (C1) should be avoided because it is more resource-intensive. PoAu (C4) algorithm should also be avoided for ET-2 application because it requires trusted nodes in the network and dilutes the trustless feature of blockchain. Both the identity management schemes may be used for ET-1 and ET-2. Similarly, for both the applications, the choice between deterministic and non-deterministic smart contracts (T1,T2) can be made based on the availability of information inside or outside the network for the execution of smart contracts. With this information, the required ingredients to build the best blockchain for P2P energy trading scenario depicted in Figure~\ref{P2P:app} can be easily identified. \\
\noindent \underline{\textbf{GI domain}}:  
Privacy requirements of green certificates GI-1 applications are less stringent. Suitable blockchain technology solutions for this application are the same as we identified for ET-1 application. 
 For behavior incentives GI-2 application with less stringent latency and throughput requirements, the choice of network, data, identity and automation management is the same as identified for SI-1 application. However, for consensus, voting-based (C3) technique can be used if there are limited number of nodes in the network and and PoAu (C4) techniques can also be adopted to conserve resources.   \\ 
\noindent \underline{\textbf{EM domain}}: 
For contract-based DRM EM-1 application, suitable technology options are the same as identified for SI-1 application. However, for real-time DRM EM-2 application, due to extremely low latency requirements, only PoAu (C4) algorithm is more suitable choice, while all other options remain the same as identified for EM-1. For distributed energy management EM-3 application, suitable technology options for network, data, identity, and automation management are the same as we identified for ET-2 application. However, for this application, due to relatively low latency requirements, PoS (C2) and PoAu (C4) techniques are more suitable for consensus management. Finally, suitable technology options for EV EM-4 application are the same as identified for ET-2 application except that for EM-4 we can also use PoAu (C4) algorithm to conserve resources. 


\begin{table*}[ht]
\centering
\fontsize{6.7pt}{6.7pt}\selectfont
\caption{Blockchain Customization for Various Smart Energy Applications.}
\begin{tabular}{|c|M{2cm}|M{4.0cm}|M{8.5cm}|} \hline 
\rowcolor[HTML]{9B9B9B}
\textbf{Domain} & \textbf{Application}  & \textbf{Blockchain Technology Options} & \textbf{Remarks} \\ \hline

\multicolumn{1}{|c|}{\multirow{3}{*}{\textbf{SI}}}
& \textbf{AMI (SI-1)} & \textbf{(N2,N3), (D2), (C2,C4), (S2), (T1)} & Blockchain integration is more suitable in new smart meter roll out programs. \\\cline{2-4} 
& \textbf{Asset management (SI-2)} & \textbf{(N2,N3), (D2), (C2,C4), (S1,S2), (T1)} & Blockchain integration is more suitable to track shared or crowd-funded RES assets. \\\cline{2-4}  
& \textbf{Grid monitoring (SI-3)} & \textbf{(N2,N3), (D2), (C4), (S2), (T1)}  & Latency requirements are few ms, even PoAu (C4) might not be able to fulfill these requirements. \\ \hline 

\multicolumn{1}{|c|}{\multirow{2}{*}{\textbf{ET}}}
& \textbf{Wholesale (ET-1)} & \textbf{(N1,N2), (D2), (C1,C2,C3,C4), (S1,S2), (T1,T2)} & Blockchain integration can eliminate existing brokers but initial implementation costs could be high. \\ \cline{2-4}  
& \textbf{P2P (ET-2)} & \textbf{(N2), (D2), (C2,C3), (S1,S2), (T1,T2)} & High potential of blockchain integration due to localized nature and lack of P2P trading platforms. \\ \hline 

\multicolumn{1}{|c|}{\multirow{2}{*}{\textbf{GI}}} 
& \textbf{Certificates (GI-1)} & \textbf{(N1,N2), (D2), (C1,C2,C3,C4), (S1,S2), (T1)}  &  Blockchain integration can eliminate existing brokers but initial implementation costs could be high. \\\cline{2-4} 
& \textbf{Incentives (GI-2)} & \textbf{(N2,N3), (D2), (C3,C4), (S2), (T1)}  & N3 can also be established if all users belong to same utility company. \\ \hline 

\multicolumn{1}{|c|}{\multirow{5}{*}{\textbf{EM}}}
& \textbf{DRM (contract) (EM-1)} & \textbf{(N2,N3), (D2), (C2,C4), (S2), (T1)}  & C2 to be avoided if stake of node is low, while S2 is listed due to KYC requirements.  \\\cline{2-4} 
& \textbf{DRM (real-time) (EM-2)} & \textbf{(N2,N3), (D2), (C4), (S2), (T1)} & Latency requirements are few sec, even PoAu (C4) might not be able to fulfill these requirements. \\\cline{2-4} 
& \textbf{Distributed EM  (EM-3)} & \textbf{(N2), (D2), (C2,C4), (S1,S2), (T1,T2)}  & T2 if external weather information is needed. \\\cline{2-4} 
& \textbf{EV (EM-4)} & \textbf{(N2), (D2), (C2,C3,C4), (S1,S2), (T1,T2)}  & C2 or C3 can also be used if they reduce costs. High potential of blockchain integration. \\ \hline 

\end{tabular}
\label{sg:app:blk}
\end{table*}

\section{Blockchain Technology Gaps for Smart Energy Systems}
\label{block:gap}
Blockchain is still evolving and there are several technology gaps, which could limit its adaptation in smart energy systems. Below we discuss some blockchain technology gaps for smart energy systems.

\noindent \underline{\textbf{Network management}}: Management of blockchain network requires appropriate protocols and algorithms. These protocols are required for transaction forwarding, data dissemination, node discovery, maintaining a list of misbehaving nodes, and limit on number of peer connections. The performance of these protocols has a direct impact on latency, throughput, and speed of transaction processing. In this context, there is a need to develop delay-aware, security-aware, privacy-aware, and scalable network management protocols for blockchain integration in smart energy systems. Moreover, the protocols must also provide flexible parameters in order to achieve various tradeoffs according to latency and throughput requirements of smart energy applications.

\noindent \underline{\textbf{Data management}}: 
Implementation of off-chain data management, which is mostly required for resource constrained nodes in smart energy systems is more challenging as it requires synchronization and availability of conventional databases. In this context, determination of optimal amount of data that should be kept on-chain and off-chain for various applications is important. Storage of off-chain data in tamper-proof manner is also challenging. Furthermore, data models and database schema can also vary across different organizations or applications. Novel techniques for handling multiple types of data models, database schema, and query processing on blockchain are also required.     

\noindent \underline{\textbf{Consensus management}}: PoAu algorithm is the fastest consensus management algorithm. However, the latency and throughput requirements of some applications are extremely stringent (in ms), and even PoAu may fail to fulfill those requirements. There is a clear need for further improvements in the consensus management techniques for smart energy applications. For example, the use of implicit consensus proposed in \cite{ren2017implicit} maybe explored. 

\noindent \underline{\textbf{Identity management}}: In several smart energy applications, due to KYC requirements enforced by the regulator, decentralized-trusted identity scheme has to be used. This scheme has less advantages as compared to more private self-sovereign identity management scheme. Recovering compromised identities can also become a challenge in some smart energy systems particularly for nodes with private or critical data.   

\noindent \underline{\textbf{Automation management}}: Security of smart contract is critical because if a smart contract is not well-written and secure, it may be hacked or invoked under different circumstances that may not represent the actual intention of the original programmer. Non-deterministic smart contract management presents even a bigger security challenge. Smart energy applications involving critical data and industrial infrastructure necessitate appropriate programs and templates for the development of secure
and well-written smart contracts. Smart contract execution often require sequential processing, which can slow down transaction verifications. Development of appropriate sharding techniques for parallel processing is therefore required to match the high performance demands of various applications. 

\noindent \underline{\textbf{Lack of suitable implementation platforms}}: 
Many popular blockchain platforms are non-modular and they do not embed appropriate technology solutions for smart energy systems. For example, the platforms lack support for off-chain data management and non-deterministic smart contracts, which are mostly required for resource constrained nodes. Therefore, development of open source and modular blockchain platforms with appropriate embedded technologies to support multiple smart energy applications is critically needed.

\section{Conclusion}
\label{con:clusion}
Blockchain technology is novel but complicated, and its integration in any domain requires the convergence of appropriate building blocks to achieve the respective desired objectives. Existing blockchain integration efforts in smart energy systems mostly use open-source blockchain platforms with embedded functionalities. These platforms are not entirely designed for energy applications and the development of blockchain based energy projects through these platforms may not provide the expected blockchain integration benefits. In this paper, we adopted a systematic approach, where we first collected the requirements of smart energy systems. After detailing the requirements for each smart energy domain, we determined the most suitable blockchain building blocks for respective smart energy systems. Accordingly, we identified blockchain technologies that meet these requirements. We further customized blockchain technologies for various smart energy applications in SI, ET, GI and EM domains. The analysis in this paper can help in the design of flexible blockchain platforms customized for smart energy systems, as well as reaping the most benefits out of blockchain integration in smart energy systems. Significant new research in blockchain technologies is still required to meet the diverse and often stringent latency, privacy, and security requirements of smart energy applications. Moreover, modular blockchain platforms, where embedded technology options can be changed on demand, would also be required to support and accelerate blockchain integration in a wide variety of smart energy applications.


\bibliographystyle{ieeetran}   
\bibliography{blockbib1short} 

\end{document}